\def \be {\begin{equation}}
\def \ee {\end{equation}}
\def \d {\widehat{\nabla}}
\def \t {\textstyle}
\documentstyle{article}
\title{LIMITS ON ANISOTROPY AND INHOMOGENEITY
FROM THE COSMIC BACKGROUND RADIATION}
\author{Roy Maartens\,\thanks{School of Maths, Portsmouth University,
PO1 2EG England}\, \thanks{Computational and Applied Maths Dept,
Witwatersrand University, 2050 South Africa} \and George FR Ellis
\,\thanks{Applied Maths Dept, University of Cape Town, 7700 South
Africa} \and William R Stoeger\,${}^\ddagger$
\thanks{Vatican Observatory Research
Group, University of Arizona, AR85721 USA}}
\begin{document}
\maketitle
\begin{abstract}
We consider directly the equations by which matter imposes anisotropies
on freely propagating background radiation, leading to a new way of
using anisotropy measurements to limit the deviations of the Universe from
a Friedmann-Robertson-Walker (FRW) geometry. This approach is complementary
to the usual Sachs-Wolfe approach: the limits obtained are not as detailed,
but they are more model-independent.
We also give new results about combined matter-radiation perturbations in an
almost-FRW universe, and a new exact solution of the linearised equations. 
\end{abstract}
\newpage
\section{Introduction}

The Cosmic Background Radiation (CBR) is the keystone of modern
cosmological analysis, in particular through use of the results of
COBE [1,2] and other [3,4] measurements of anisotropies in the CBR
to help us understand the nature of inhomogeneities in the universe
(see [5] for a review of these observations).
This paper presents a way of analysing this relationship that is an
alternative to the usual analyses based largely on the Sachs-Wolfe
effect (modified by astrophysical effects).
Our analysis proceeds from slightly different assumptions than those
usually made (though essentially compatible with them); it is to a
considerable degree more model-independent than they are.\\

In [6] we established a theoretical framework for investigating the
direct implications of CBR anisotropies (i.e. those that follow
without assuming particular inflationary or other evolutionary models
for the universe). In that paper, we set up fully covariant and
gauge invariant evolution and constraint equations governing the
perturbations of the photon distribution and metric. These equations
were then used to show that if
\begin{quotation}
{\bf A1}: all fundamental observers measure the CBR to be almost isotropic
in some domain,
\end{quotation}
then it follows that
\begin{quotation}
{\bf A2}: the spacetime geometry is almost Friedmann-Robertson-Walker
(`FRW') - i.e. the shear, vorticity, spatial gradients and Weyl tensor
are almost zero, and the metric may be put into perturbed-FRW form
in that domain.
\end{quotation}
The latter is the assumption which underlies the usual Sachs-Wolfe
analyses - for it is the starting point used to set up the Sachs-Wolfe
equations.\\

In this paper, we use the formalism of [6], extending its results to
find quantitative limits on the anisotropy and inhomogeneity of spacetime
set by anisotropies in the CBR, {\it without} assuming a particular model
for the origin of such perturbations. We end up with a series of estimates
(Section 4) relating the inhomogeneity and anisotropy of the universe
directly to the background radiation anisotropies.
In addition, we find some new exact results on perturbations in
almost--FRW universes with both matter and radiation (Sections 2 and 5).
For a class of almost flat--FRW universes, we reduce the full set of
linearised dynamical equations to two linear ODE's, and give a new exact
solution at late times (Section 5).\\

The difference from the more usual
analyses is that they consider observations of the CBR made
at one space time event $P$ (`here and now'), relating them to assumed
perturbations of the universe at the time of decoupling, these in turn
taken to arise from some particular evolutionary history or other. Here we
make no such evolutionary assumptions; however (cf. {\bf A1}) we assume
the nature of CBR anisotropies is known not only at our own space-time
position $P$ but also throughout an open domain containing that event
(eventually chosen to represent the period between last scattering and the
present, in a region containing our world-line and past light cone). \\

At first glance this seems to
make our analysis far more dependent on unobservable data than the standard
approach. However this is an illusion, for that approach builds in equivalent
assumptions at the outset, but in a rather hidden way, because it assumes
{\bf A2}, which cannot be {\it proved} on the basis of observational
evidence alone [7,8]. It can only be {\it deduced} from such data on the
basis of some kind of Copernican assumption such as {\bf A1}, which is
{\it not} directly testable [6-8]. Our approach helps make clear precisely
what these hidden assumptions are, and thereby enables one to test how weak
they can be and still allow deduction of the desired results. In fact,
as pointed out by George Smoot (private communication), the Copernican
assumption is in principle partially testable. This follows since the
Sunyaev--Zeldovich effect allows us to confirm to some degree that
distant galaxies do see nearly isotropic CBR, for otherwise the
scattered radiation would have a significantly distorted black--body
spectrum.\\

In the covariant approach we work in the real non-FRW spacetime, rather
than starting from an assumed FRW model and perturbing away from it.
The former approach considers the full set of dynamical equations that
govern the real variables. The latter approach risks missing certain
effects [9,10] or masking underlying
assumptions. We provide an example of the second kind by showing that
the Boltzmann equation imposes constraints on the photon distribution
if the monopole moment is assumed to be Planckian to first order.
The covariant formalism for analysing fluid inhomogeneities [9,11,12] is a
development of Hawking's approach [13], and gives a gauge invariant
alternative to Bardeen's formalism [14]. The covariant approach to the photon
distribution function in this paper and [6] is based on [15,16], and is
an alternative to the application of a Bardeen-type formalism, as
presented in [17,18] (in different contexts from that of this paper). \\

Despite the success of the standard inflationary models with dark matter and
critical density [19,20], current CBR observations are
consistent with alternative models, and do not by themselves
give independent tests of inflation [21,22]. Furthermore, future observations
could produce problems for the standard models.
The covariant approach provides a clear and direct relation between
observational and theoretical quantities, unobscured by particular
gauge choices or by the complexities of harmonically determined
variables. Furthermore, we do not impose any specific model to
generate fluctuations in the CBR. Thus we investigate, as far
as possible, what is determined directly by observations of the CBR made
by the family of fundamental observers.
Where we are forced to make additional assumptions, they are made
about observational quantities - and are thus in principle falsifiable
by observation, provided we make some kind of Copernican assumption
such as {\bf A1}, stating that all fundamental observers see the same kind of
thing (the nature of the required assumptions is clarified below).
In this sense, we provide a framework for comparing
and testing various models, in which there is as clear as possible a
distinction between observed and assumed properties.
\newline

\noindent{\bf Notation:}\newline
The metric $g_{ab}$ has signature $(-,+,+,+)$. Einstein's
gravitational
constant, the speed of light in vacuum, and Planck's constant are
one. Round brackets on indices denote symmetrisation, square
brackets anti-symmetrisation.
$\nabla_a$ is the covariant derivative defined by $g_{ab}$. Given a
four--velocity $u^a$, the associated projection tensor is
$h_{ab} = g_{ab} +u_a u_b$, and the comoving time derivative and
spatial gradient are
$$
\dot{Q}_{a...b} \equiv u^c \nabla_c Q_{a...b}\,,
$$
$$
\d_c Q_{a...b} \equiv h_c{}^d h_a{}^e \dots
h_b{}^f \nabla_d Q_{e...f}
$$
for any tensor $Q_{a...b}$ (in [6] we used ${}^3\nabla_a$ for
$\d_a$). If the tensor is spatial, we define
$$
\mid Q_{a...b} \mid \equiv (Q_{a...b} Q^{a...b})^{1/2}\,.
$$

Given a smallness parameter $\epsilon$, $O[N]$ denotes $O(\epsilon^N)$
and $A \simeq B$ means $A - B = O[2]$ (i.e. these variables are equivalent
to $O[1]$). When $A \simeq 0$ we shall regard $A$ as vanishing (for it is
zero to the accuracy of the first-order calculations that are the concern
of this paper).

\section{Covariant and Gauge-Invariant Analysis}

The fundamental observers are assumed comoving with the cosmological matter,
which is modelled by dust with mass--energy density $\rho$,
and which is non--interacting with the radiation (as we are considering
the epoch after last scattering). The physically preferred four--velocity
$u^a$ of this matter is a suitable average over peculiar velocities (which are
small). The matter flow is characterised by $u^a$ and its rate of expansion
$\Theta$ ($=3H=3\dot{S}/S>0$, where $H$ is the Hubble parameter and $S$ the
scale factor), shear $\sigma_{ab}$, and vorticity
$\omega_{ab}$, all non-zero in general; however the flow lines are
geodesic: $\dot{u}^a = 0$ (consequent on the vanishing of the matter
pressure).\\

The frame defined by $u^a$ defines an invariant 3+1 splitting of tensors
[23]. In particular, for a photon four-momentum
\be
p^a = E(u^a + e^a)\,, ~~~~~ e_a u^a = 0\,, ~~~~~ e_a e^a = 1\,,
\ee
where $E$ is the photon energy and $e_a$ the direction of photon
momentum, relative to fundamental observers. Then the CBR distribution
function may be expanded as [15]
\be
f(x^c,E,e^d) = F(x^c,E)+F_a(x^c,E)e^a+F_{ab}(x^c,E)e^a e^b+ \dots
\ee
where the covariant harmonics (multipole moments) $F_{a_1 \dots a_L}(x^c,E)$
for $L \ge 1$ are symmetric trace-free tensors orthogonal to $u^a$, that
provide a measure of the deviation of $f$ from exact isotropy (as measured
by $u^a$). If the CBR is almost isotropic after last scattering for all
fundamental observers, then [6]:
\be
F,~\dot{F} = O[0] \, ,~~~~~~~ F_{a_1 \dots a_L},~\nabla_b
F_{a_1 \dots a_L} = O[1] \, ,~~~~~ L \ge 1 \, .
\ee

Energy integrals of the first three harmonics
define the radiation energy density, energy flux and anisotropic
stress [15]:
\be
\mu = 4\pi \int_0^\infty E^3 F dE = 3p = O[0] \,,
\ee
\be
q_a = {{4\pi}\over 3} \int_0^\infty E^3 F_a dE = O[1]\,,
\ee
\be
\pi_{ab} = {{8\pi}\over 15}\int_0^\infty E^3 F_{ab} dE = O[1]
\ee
(in [6] we used $\mu_R$ for $\mu$). We will also need the integral of
the third harmonic:
\be
\xi_{abc} = {{8\pi}\over 35}\int_0^\infty E^3 F_{abc} dE = O[1]\,.
\ee
Note that the total energy--momentum tensor is made up of matter and
radiation contributions:
$$
T_{ab} = (\rho + \mu)u_a u_b +{\t{1\over 3}}\mu h_{ab}+
\pi_{ab}+2u_{(a}q_{b)}\, .
$$

\subsection{Covariant linearised harmonics
of the Boltzmann \newline equation}

The Boltzmann equation in curved spacetime
$$
p^a \left( {\partial \over \partial x^a} - \Gamma^c{}_{ab} p^b
{\partial \over \partial p^c} \right) f(x^d,p^e) = C[f]
$$
may be decomposed into covariant harmonic equations via (1), (2).
The full (exact and non-linear) results are
given in [16, p501]. For the collision--free and
zero acceleration case, the linearised zero, first and second harmonic
equations are:
\be
E \dot{F}-{\t{1 \over 3}}E^2 \Theta {\partial F \over
\partial E} +
{\t{1 \over 3}}E \d_a F^a \simeq 0\,,
\ee
\be
E \dot{F}_a - {\t{1\over 3}}E^2 \Theta {\partial F_a \over
\partial E} +
E \d_a F + {\t{2\over 5}} E \d_b
F^b{}_a \simeq 0\,,
\ee
\be
E \dot{F}_{ab}-{\t{1\over 3}}E^2 \Theta {\partial F_{ab} \over
\partial E} - E^2 \sigma_{ab} {\partial F \over \partial E}+E \widehat
{\nabla}_{(a}F_{b)} -
\ee
$$
-{\t{1\over 3}}h_{ab}\d_c F^c+
{\t{3\over 7}}E \d_c F^c{}_{ab} \simeq 0\,.
$$

These are the fundamental (covariant) equations governing the dynamics of
radiation at a microscopic level.

\subsection{Covariant linearised evolution and constraint
\newline equations}
If (8--10) are multiplied by $E^2$ and
integrated over all photon energies, then they produce the radiation
conservation equations governing $\mu$ and $q_a$, as well as the
crucial evolution equation for $\pi_{ab}$ (given for the first time in
[6], in full non--linear form). These equations and the remaining
(linearised) conservation, Einstein, Ricci and Bianchi equations are
as follows, obtained from [6] and the general exact equations of [23]
(with $\dot {u}_a=0$ but allowing for an imperfect energy-momentum tensor).\\

\noindent a) Matter and radiation energy and momentum conservation:
\be
\dot{\rho} + \Theta \rho = 0 \, ,~~~~~~~~ \dot{u}_a = 0\, ,
\ee
\be
\dot{\mu} + {\t{4\over 3}}\Theta \mu + \d_a q^a
\simeq 0 \, ,
\ee
\be
\dot{q}_a + {\t{4\over 3}}\Theta q_a +{\t{1\over 3}}
\d_a \mu +
\d_b \pi^b{}_a \simeq 0 \, .
\ee
b) Evolution of radiation anisotropic stress tensor:
\be
\dot{\pi}_{ab}+{\t{4\over 3}}\Theta \pi_{ab}+
{\t{8\over 15}}\mu \sigma_{ab} +
2\d_{(a} q_{b)} -
{\t{2\over 3}}h_{ab} \d_c
q^c +\d_c \xi^c{}_{ab} \simeq 0 \, .
\ee
c) Einstein, Ricci and Bianchi propagation equations:
\be
\dot{\Theta} + {\t{1\over 3}}\Theta^2 + \mu +
{\t{1\over 2}}\rho \simeq 0 \, ,
\ee
\be
\dot{\sigma}_{ab} + {\t{2\over 3}}\Theta \sigma_{ab} +
E_{ab} - {\t{1\over 2}}\pi_{ab} \simeq 0 \, ,
\ee
\be
\dot{\omega}_{ab} + {\t{2\over 3}}\Theta \omega_{ab}
\simeq 0  \, ,
\ee
\be
\dot{E}_{ab} + \Theta E_{ab} + \d^d H_{(a}{}{}^c
\varepsilon_{b)cd} +({\t{1\over 2}}\rho +
{\t{2\over 3}}\mu) \sigma_{ab} \, +
\ee
$$
+\, {\t{1\over 2}}\dot{\pi}_{ab}
+{\t{1\over 6}}\Theta \pi_{ab} +
{\t{1\over 2}}\d_{(a} q_{b)} -
{\t{1\over 6}}h_{ab} \d_c q^c
\simeq 0 \, ,
$$
\be
\dot{H}_{ab}+\Theta H_{ab}-\d^d E_{(a}{}{}^c
\varepsilon_{b)cd}+
{\t{1\over 2}}\d^d \pi_{(a}{}{}^c
\varepsilon_{b)cd} \simeq 0 \, .
\ee
d) Einstein, Ricci and Bianchi constraint equations:
\be
q_a - {\t{2\over 3}}\d_a \Theta +
\d^b
(\sigma_{ba}+\omega_{ba}) = 0 \, ,
\ee
\be
\d_a \omega^a = 0 \, ,
\ee
\be
H_{ab} +\d^d [\sigma_{(a}{}{}^c +\omega_{(a}{}{}^c]
\varepsilon_{b)cd} \simeq 0 \, ,
\ee
\be
\d_b E^b{}_a -
{\t{1\over 3}}\d_a(\mu+\rho)+
{\t{1\over 2}}\d_b \pi^b{}_a +
{\t{1\over 3}}\Theta q_a \simeq 0 \, ,
\ee
\be
\d_b H^b{}_a - (\rho +
{\t{4\over 3}}\mu)\omega_a +
{\t{1\over 2}}\varepsilon_{abc} \d^b q^c
\simeq 0 \, ,
\ee
where $E_{ab}$, $H_{ab}$ are the electric and magnetic parts of the
Weyl tensor, and $\varepsilon_{abc} \equiv \eta_{abcd} u^d$,
with $\eta_{abcd}$ the spacetime permutation tensor.\\

It follows [6] from these equations and (3--7) that:
\be
\psi,~\dot{\psi} = O[0] \, ,~~~ \d_a \psi = O[1] \, ,~~~
\sigma_{ab},~\omega_{ab},~E_{ab},~H_{ab} = O[1] \, .
\ee
where $\psi \equiv \mu,~\Theta,~\rho$.
These qualitative results from [6] will be made more
quantitative in section 4.

\subsection{Integrability conditions and conserved quantities}
In the case of zero acceleration, the linearised form of the identity
[11] governing the commutation of time and spatial derivatives is:
\be
\d_a \dot{\Psi} \simeq (\d_a \Psi)^
{\displaystyle\cdot}
+{\t{1\over3}}\Theta \d_a\Psi
\ee
where $\Psi$ is any $O[1]$ spatial tensor or any scalar with $O[1]$
gradient. The commutation of spatial derivatives themselves is given
by the projected Ricci identities [11], which imply the exact identity
\be
\d_{[a}\d_{b]}\psi = -\dot{\psi}\omega_{ab}
\ee
where $\psi$ is any scalar, and for a nearly-FRW
spacetime, the linearised conditions
\be
\d_{[a}\d_{b]}\psi_c \simeq
{k \over S^2}h_{c[a}\psi_{b]}
\ee
where $\psi_a$ is any $O[1]$ spatial vector, and $k=0,1,-1$ is the
spatial curvature index of the limiting (background) spacetime; and
\be
\d_{[a}\d_{b]}\psi_{cd} \simeq
-{k \over S^2}(h_{c[a}\psi_{b]d}-\psi_{c[a}h_{b]d})
\ee
where $\psi_{ab}$ is any $O[1]$ spatial tensor. These identities could
be overlooked in an approach that starts from an FRW background
solution and perturbs away from it. The
integrability conditions implicit in (27--29) lead to the new results:
\begin{quotation}
\noindent {\bf (I)} {\it if the covariant vector perturbations are
spatially homogeneous to first order, then the vorticity vanishes to first
order (i.e. non-zero terms are at most second order), and either all vector
perturbations vanish to first order, or the spacetime has a flat
FRW background.} \\
\newline {\bf (II)} {\it if the covariant tensor perturbations are
spatially homogeneous to first order, then the spacetime is
either FRW to first order, or it has a flat FRW background.}
\end{quotation}

The first result follows from (27), which shows that
$\omega_{ab}\simeq 0$ since $\d_a\psi$ is a vector
perturbation for $\psi=\mu,\rho$, and from (28), which implies
$kS^{-2}\psi_a \simeq 0$ for $\psi_a=q_a$ or any other vector
perturbation. The second result follows from (28) and (29), which
imply
$$
kS^{-2}\psi_a \simeq 0 \simeq kS^{-2}\psi_{ab}
$$
for any vector perturbation $\psi_a$ (since $\d_a\psi_b$
is a tensor) or tensor perturbation
$\psi_{ab}$. Thus either $k/S^2 \simeq 0$
(the background FRW model is flat to the accuracy we are working), or
$$
\d_a\mu \simeq 0 \simeq \d_a\rho\, ,~~~~
q_a \simeq 0\,,~~~~\pi_{ab} \simeq 0\,,~~~~\d^c\xi_{cab}
\simeq 0\,.
$$
In the latter case, (27) implies $\omega_{ab} \simeq 0$, (14) implies
$\sigma_{ab} \simeq 0$, and then (20) gives $\d_a\Theta
\simeq 0$, while (16) and (22) give $E_{ab} \simeq 0
\simeq H_{ab}$. Thus the spacetime is FRW to first order (i.e. it
differs from FRW at most by second order terms). \hfill $\Box$\\

We can derive further linearised integrability results
        that hold in general (i.e. without assuming homogeneity of
        vector or tensor perturbations),
by covariant differentiation of the dynamical equations. For example,
taking the gradient of (24), and using (21) and
$\d_{[a}\d_b q_{c]}\simeq 0$ (which
follows from (28)), we get
\be
\d_a\d_b H^{ab} \simeq 0 \, .
\ee
Similarly, using the contraction of (29) for $\psi_{cd}= \omega_{cd}$,
the gradient of (20) gives
\be
\d_a\d_b\sigma^{ab}+\d_aq^a
-{\t{2\over3}}\d^2\Theta \simeq 0 \, ,
\ee
while (23) yields
\be
\d_a\d_b(E^{ab}+{\t{1\over2}}
\pi^{ab})+{\t{1\over3}}\Theta\d_aq^a
-{\t{1\over3}}\d^2(\mu+\rho) \simeq 0 \, .
\ee

Finally, we note the existence of various quantities that are
conserved to first order along the matter flow. For example, (17)
immediately shows that
$$
(S^2 \omega_{ab})^{\displaystyle\cdot} \simeq 0 \, ,
$$
while (26), (19) and (16) imply that
\be
\dot{H}^*_{ab}+\Theta H^*_{ab} \simeq 0\, ,~~~~~~ H^*_{ab} \equiv H_{ab}+
\d^d\sigma_{(a}{}{}^c\varepsilon_{b)cd}\, .
\ee
Then (33) and (11) give
$$
\left({{H_{ab}+\d^d\sigma_{(a}{}{}^c\varepsilon_{b)cd}}\over
\rho}\right)^{\displaystyle\cdot} \simeq 0 \, .
$$
It therefore follows that if either the vorticity or $H^*_{ab}/\rho$ are
negligible (i.e. $O[2]$) at last scattering, they remain so at all
subsequent times.

\section{Temperature Anisotropy of the CBR}

It is important to realise that the covariant dipole moment $F_a$ of
the CBR distribution (see (2)), although dependent upon the choice of
$u^a$, cannot be set to zero by this choice, since
it is frequency--dependent and its vanishing implies special
conditions on the anisotropy of $f$.
Since the $u^a$--frame is physically defined by the matter, and is
already assumed to have been corrected for local peculiar velocities,
$F_a$ represents a possible residual intrinsic frequency dependent dipole
moment of the CBR distribution relative to the matter, with invariant
significance. This distributional dipole however contains much
more information than the dipole of
CBR {\it temperature} anisotropy, which is in fact proportional to the
energy flux $q_a$ given by (5) (see (41) below).\\

Even for the temperature dipole however,
one cannot separate the intrinsic dipole from that induced by peculiar
velocity of the observer [19,24]. It is standard to assume that the
intrinsic temperature dipole is negligible after correction for
peculiar velocity. This is equivalent to the non--trivial assumption
that the average
four--velocity of the matter coincides with the energy--frame [12]
four--velocity of radiation. Although it can be justified for
adiabatic perturbations within the standard model [24], we will not
make this special assumption, so that we allow for an intrinsic dipole
in the temperature (i.e. non--zero $q_a$) after correction for local
peculiar velocities. This approach accommodates future improvement
of observational results for the peculiar velocity which are
{\it independent} of the CBR observations, and which may reveal a
non--negligible residual temperature dipole. From this point of
view, the current limits on the intrinsic dipole should be related to
the current uncertainties in the local peculiar velocity.\\

Another aspect of the dipole moment $F_a$ which appears not to have
been previously recognised is its link to deviations from a thermal
Planck spectrum in the monopole moment $F$. This aspect is hidden if one
starts from a background FRW solution and then perturbs - rather than
considering the real non--FRW solution. The latter approach shows via
the Boltzmann equation that non--trivial constraints are imposed on
the dipole if $F$ is Planckian to first order (as strongly indicated by
observations [5]). For suppose that
\be
F(x^a,E) \simeq 2 \left[\exp\left( {{E}\over{kT(x^a)}} \right)-1 \right]^{-1}
\ee
where $k$ is Boltzmann's constant. Then (34) and the Boltzmann
monpole harmonic equation (8) imply
\be
\d_a F^a \simeq \left({\dot{T}\over T}+{\dot{S}\over S}
\right) {{3E/kT}\over {1-\cosh(E/kT)}}\, .
\ee
Thus the dipole moment is subject to the restriction (35) if the monopole
moment is Planckian to $O[1]$. \\

One consequence of (35) is
\be
{{\dot{T}}\over T} = -{{\dot{S}}\over S} +O[1] \, .
\ee
In contrast to many other treatments (where $TS=~$constant),
$T$ is {\it not} a fictitious background temperature, but is the
gauge-invariant average temperature in the actual spacetime.
Now observations of the CBR measure temperatures in different directions
on the sky. The full--sky average temperature $T(x^a)$ at event $x^a$
is determined by the monopole harmonic of the photon distribution:
\be
\mu(x^a) = a [T(x^a)]^4 = 4\pi \int_0^\infty E^3 F(x^a,E) dE
\ee
on using (4), where $a$ is the Stefan--Boltzmann constant.
A directional temperature is determined by {\it all} the harmonics (2)
via the directional energy density per unit solid angle that
is defined by the integrated (bolometric) brightness [25] (see also
[26--28])
\be
I(x^a,e^b) = \int_0^\infty E^3 f(x^a,E,e^b) dE =
{a \over {4\pi}} [T(x^a)+\delta T(x^a,e^b)]^4
\ee
which defines the gauge-invariant fluctuation $\delta T(x^a,e^b)$.\\

The covariant multipole moments $\tau_{a_1 \dots a_L}(x^b)$
$(L \ge 1)$ of temperature anisotropy are trace-free, symmetric
tensors orthogonal to $u^a$, defined by
\be
\tau \equiv {{\delta T}\over T} = \tau_a e^a +
\tau_{ab} e^a e^b + \tau_{abc} e^a e^b e^c + \dots
\ee
By (38), (37), (2) and (4) they are given in general, to a good
approximation, as normalised
integrals of the covariant distribution harmonics:
\be
\tau_{a_1 \dots a_L} \simeq \left({1\over4\mu}\right)
4\pi \int_0^\infty E^3 F_{a_1 \dots a_L} dE \,.
\ee
These moments give a covariant and gauge--invariant description of
the CBR temperature variation, with spectral information integrated
out. By (5--7), (40) gives the dipole, quadrupole and octopole as:
\be
\tau_a\simeq {3q_a \over 4\mu}\,,~~~~~~~\tau_{ab} \simeq {15\pi_{ab}
\over 8\mu}\,,~~~~~~~\tau_{abc} \simeq {35\xi_{abc} \over 8\mu}\,.
\ee

The $\tau_{a_1 \dots a_L}$ are a frame-independent alternative to the
usual multipole
coefficients $A_{LM}$ in an expansion in spherical harmonics $Y_{LM}$.
If we choose a standard triad in the rest space of $u^a$ such that
$e^a=(0,\sin \theta \cos \phi,\sin \theta \sin\phi,\cos \theta)$, then
the two approaches are linked (cf. [15]) by
$$
\tau(x,{\bf e})=\sum_{L=1}^\infty \sum_{M=-L}^L A_{LM}(x) Y_{LM}
(\theta,\phi) = \sum_{L=1}^\infty \tau_{a_1 \dots a_L}(x)
e^{a_1} \dots e^{a_L}\,.
$$
The correlation function
$$
C(\alpha)= \langle\tau(x,{\bf e})\tau(x,{\bf e'})\rangle~,~~~~
e^a e'_a=\cos \alpha \,,
$$
where $\langle\dots\rangle$ denotes a statistical average, is the key
quantity in actual observations.
In this paper we will not consider the detailed
statistical analysis of the correlation function, which is
given for example in [19--21,24,26,29--31], where an
inflationary model for perturbations is assumed. Our concern here
is with
the underlying principles of how to relate observational limits to
properties of the spacetime geometry in a model-independent way.  \\

Current CBR observations place limits on $\tau_{a \dots b}(t_0,
{\bf y}|_C)$ where $t_0$ is the proper time along our worldline $C$
since last scattering and ${\bf y}|_C$ are comoving coordinates of
$C$. By {\bf A1}, these limits may be extended to hold
on each worldline ${\bf y}$ at a proper time $t_0$ along the worldline
after last scattering:
$$
\mid \tau_{a_1 \dots a_L}(t_0,{\bf y})\mid < \epsilon_L (t_0)\,.
$$
As in [6], we assume that anisotropies are $O[1]$ back to last
scattering, and extend the limit to hold for all times $0<t \le t_0$,
thus obtaining the assumption
\begin{quotation}
{\bf B1}: there exist $O[1]$
constants $\epsilon_L$ such that $\epsilon_L(t) \le \epsilon_L$.
Hence for any event after last scattering
\end{quotation}
\be
\mid \tau_{a_1 \dots a_L} \mid < \epsilon_L \,.
\ee
In principle (and possibly not too far off in practice), observations
place limits on the comoving time derivatives of the multipoles.
As before we assume
\begin{quotation}
{\bf B2}: there exist $O[1]$ constants $\epsilon^*_L, \epsilon^{**},
\epsilon^{***}$ such that
\end{quotation}
\be
{\mid \dot{\tau}_{a_1 \dots a_L}\mid}
<  \epsilon^*_L \, \Theta\,,~~|\ddot{\tau}_{a_1 \dots a_L}|<
\epsilon^{**}_L \Theta^2~,~~|\ddot{\tau}^{\displaystyle\cdot}_
{a_1 \dots a_L}|<\epsilon^{***}_L \Theta^3~,
\ee
where we have normalised the derivatives relative to the expansion
of the universe (recall that $\Theta >0$ is $O[0]$), and we will not
need the higher derivative limits. Since it is
effectively impossible for us to move cosmological
distances off $C$, there will not be direct observations of the
spatial derivatives of the multipoles. However we will need limits
on spatial and mixed derivatives up to third order, and so assume
also, on general plausibility
grounds (given the basic Copernican assumption), that additionally:
\begin{quotation}
{\bf B3}: there exist $O[1]$ constants $\epsilon'_L, \epsilon''_L,
\epsilon'''_L$ such that
\end{quotation}
\be
{\mid \d_c \tau_{a_1 \dots a_L} \mid}
< \epsilon'_L \Theta~,~~|\d_d \d_c \tau_{a_1 \dots a_L}|<\epsilon''_L
\Theta^2~,~~|\d_e \d_d \d_c \tau_{a_1 \dots a_L}|<\epsilon'''_L \Theta^3~,
\ee
and constants $\epsilon'^*_L, \epsilon'^{**}_L, \epsilon''^*_L$ such
that
\be
|(\d_c \tau_{a_1 \dots a_L})^{\displaystyle\cdot}|<\epsilon'^*_L
\Theta^2~,~~|(\d_c \tau_{a_1 \dots a_L})^{\displaystyle\cdot}{}^
{\displaystyle\cdot}|<\epsilon'^{**}_L \Theta^3~,~~
|(\d_d \d_c \tau_{a_1 \dots a_L})^{\displaystyle\cdot}|< \epsilon''^*_L
\Theta^3~.
\ee

One should note that we expect all the quantities $\epsilon$ defined
here to be {\it very} small: probably at most $10^{-4}$. The exact
isotropic case considered by Ehlers, Geren, and Sachs [32] corresponds
to $\epsilon_L = \epsilon^*_L = \epsilon'_L = 0$, and is a special case
of what follows. In fact, our special case is a small generalisation
of the EGS result to include non--interacting dust matter in the
source of the gravitational field.

\section{Model-independent Limits on Spacetime \newline
Anisotropy and Inhomogeneity}
The observationally based limits (42) on temperature anisotropy lead
directly via (41) to the limits
\be
|q_a|<{\t{4\over3}}\mu\epsilon_1~,~~~~~|\pi_{ab}|<
{\t{8\over15}}\mu\epsilon_2~,~~~~~|\xi_{abc}|<
{\t{8\over35}}\mu\epsilon_3
\ee
on the radiation anisotropy tensors. Limits on the derivatives of the
radiation tensors arise from differentiating (41), using (42--45) and
the evolution and constraint equations of Section 2.2. For example,
the limits on derivatives of $q_a$ that we will need are
\be
|\dot{q}_a|<{\t{4\over3}}\mu\Theta({\t{4\over3}}
\epsilon_1+\epsilon^*_1)~,~~~
|\d_aq_b|<{\t{4\over3}}\mu\Theta\epsilon'_1~,
\ee
\be
|\ddot{q}_a|<{\t{4\over 27}}\mu\Theta^2[(20+4\Omega_R+2\Omega_M)
\epsilon_1+24\epsilon^*_1+9\epsilon^{**}_1]~,~~
|\d_a \d_b q_c|<{\t{4\over 3}}\mu\Theta^2 \epsilon''_1~,
\ee
\be
|(\d_a q_b)^{\displaystyle\cdot}|<{\t{4\over 9}}\mu\Theta^2
(3\epsilon'^*_1+4\epsilon'_1)~,~~
|(\d_a \d_b q_c)^{\displaystyle\cdot}|<{\t{4\over9}}\mu\Theta^3
(3\epsilon''^*_1+4\epsilon''_1)~,
\ee
with similar expressions for $\pi_{ab}, \xi_{abc}$. The density
parameters are
$$
\Omega_R={\mu \over 3H^2}~,~~~~\Omega_M={\rho \over 3H^2}~.
$$

Now (46) and (47--49) (with their counterparts for $\pi_{ab}$ and
$\xi_{abc}$) are used in the evoluiton and constraint equations and
their derivatives in order to find limits on the geometric and
dynamic quantities that characterise the deviation of the universe
from FRW form. We emphasise that the following limits are
gauge--invariant, covariant and imposed directly by observational
quantities, i.e. the temperature dipole, quadrupole, octopole and
their derivatives.\\

 From (13) and (14) we get immediately
\be
{{|\d_a\mu|}\over{\mu}}=4{{|\d_aT|}\over
T} < H (8
\epsilon_1+12\epsilon^*_1+{\t{72\over5}}\epsilon'_2)\,,
\ee
\be
{{|\sigma_{ab}|}\over{\Theta}} < {\t{8\over3}}\epsilon_2+
\epsilon^*_2+5\epsilon'_1+{\t{9\over7}}\epsilon'_3\,.
\ee
The remaining limits require more complicated manipulation of the
evolution and constraint equations. From (27), using $\d_b$(13) to get
$|\d_a \d_b \mu|$, we find
\be
{|\omega_{ab}| \over \Theta}<9\epsilon'_1+3\epsilon'^*_1
+{\t{6\over5}}\epsilon''_2~.
\ee
 From (23), using $\d_c$(16), $\d_c$(14) and [$\d_c$(14)]${}^
{\displaystyle\cdot}$ to get limits on $|\d_c E_{ab}|$,
$|\d_c \sigma_{ab}|$ and $|(\d_c \sigma_{ab})^{\displaystyle\cdot}|$,
we find
$$
{|\d_a \rho| \over \Theta}<{\t{27\over2}}H\epsilon'_2+
\left({\Omega_R \over \Omega_M}\right)H[12\epsilon_1+12\epsilon^*_1
+61\epsilon'_2]+
$$
\be
 + \left({3 \over \Omega_M}\right)H[165\epsilon''_1+45\epsilon''^*_1+
110\epsilon'_2+69\epsilon'^*_2+9\epsilon'^{**}_2+18\epsilon''_3]~.
\ee
 From $\d_a$(12), using (13)${}^{\displaystyle\cdot}$ and (26) to get
$|(\d_a \mu)^{\displaystyle\cdot}|$, we find
\be
{|\d_a \Theta| \over \Theta}<H[50\epsilon_1+51\epsilon^*_1+
9\epsilon^{**}_1+3\epsilon''_1+{\t{24\over5}}\epsilon^*_2+
18\epsilon'_2+{\t{18\over5}}\epsilon'^*_2]+
4(2\Omega_R+\Omega_M)H\epsilon_1~.
\ee
 From (16), using (14)${}^{\displaystyle\cdot}$ to get
$|\dot{\sigma}_{ab}|$ and (12), (15), (26), we find
\be
{|E_{ab}| \over \Theta}<H[50\epsilon'_1+15\epsilon'^*_1+{\t{88\over3}}
\epsilon_2+14\epsilon^*_2+3\epsilon^{**}_2+{\t{66\over7}}
\epsilon'_3+{\t{9\over7}}\epsilon'^*_3]+
{\t{4\over45}}(11\Omega_R+15\Omega_M)H\epsilon_2~.
\ee
 From (22), using $\d_c$(27) (with $\psi=\mu$) and $\d_c \d_b$(13)
to get $|\d_c \omega_{ab}|$, and $\d_c$(14) to get $|\d_c \sigma_{ab}|$,
we find
\be
{|H_{ab}| \over \Theta}<H[45\epsilon''_1+9\epsilon''^*_1+9\epsilon'_2
+3\epsilon'^*_2+{\t{18\over5}}\epsilon'''_2+{\t{9\over7}}
\epsilon'''_3]~.
\ee

These equations show explicitly the role of the dipole, quadrupole
and octopole. For example, (50) shows that the gradient of radiation
energy density or average temperature, which reflects
inhomogeneous deviations from FRW spacetime, is bounded
by the limits on both the dipole and quadrupole of temperature
anisotropy, while by (51), the shear,
which reflects anisotropic deviations from FRW spacetime, is bounded
by the limits on the dipole, quadrupole and octopole. \\

Note that if we
follow the usual assumption that the CBR temperature dipole is
negligible (after correction for local peculiar velocities), then by
(41), (42--45) it follows that all the $\epsilon_1$'s vanish:
$$
\tau_a \simeq 0 ~~ \Leftrightarrow ~~ q_a \simeq 0 ~~ \Leftrightarrow
{}~~ \epsilon_1 =\epsilon^*_1 =\cdots= \epsilon''^*_1 = 0 \,.
$$
In this case, the bounds in (50--56) are reduced, so that a negligible
dipole reduces the limits on anisotropy and inhomogeneity.\\

Using (12) and (37), (50) may be re-written
$$
{{|\d_aT|}\over {|\dot{T}|}}<{\t{1\over2}}
\epsilon_1+3\epsilon^*_1+{\t{18\over5}}\epsilon'_2\,,
$$
from which it follows that
$$
{{|\d_aT|}\over T} \ll {{|\dot{T}|}\over T}~~
\Leftrightarrow~~d_R \gg t_R~,
$$
where
\be
d_R={T\over{|\d_aT|}}\,, ~~~~~~~~~
t_R={T\over{|\dot{T}|}}
\ee
are characteristic length and time scales defined by the CBR. In
practice $\epsilon^*_L$,$\epsilon^{**}_L$ and especially
$\epsilon'_L$, $\epsilon''_L,\cdots$ are not known from observations.
In order to produce
more useful versions of the results, we need a reasonable estimate of
these quantities. First, we make the reasonable assumption
\begin{quotation}
{\bf C1}: the spatial gradients of the temperature multipoles are not
greater than their time derivatives:
\end{quotation}
$$
\epsilon'_L \le \epsilon^*_L~;~\epsilon''_L,\epsilon'^*_L\le
\epsilon^{**}_L~;~\epsilon'''_l,\epsilon'^{**}_L,\epsilon''^*_L \le
\epsilon^{***}_L~.
$$

Next, we can estimate $\epsilon^*_L, \epsilon^{**}_L,
\epsilon^{***}_L$ by invoking the characteristic time
$t_R$ defined by (57), leading to assumption
\begin{quotation}
{\bf C2}: the bounds on the time derivatives of the temperature
harmonics are estimated by
$\Theta\, \epsilon^*_L \simeq \epsilon_L/t_R~$, $\Theta^2
\epsilon^{**}_L \simeq \epsilon_L/t_R^2~$, $\Theta^3\epsilon^{***}_L
\simeq \epsilon_L/t_R^3~$, so that, using (36)
\end{quotation}
$$
\epsilon^*_L \simeq {\t{1\over3}}\epsilon_L~,~
\epsilon^{**}_L \simeq {\t{1\over9}}\epsilon_L~,~
\epsilon^{***}_L \simeq {\t{1\over27}}\epsilon_L~.
$$

We can now use {\bf C1--2} to re-cast (50--56) in terms of the
observationally realistic $\epsilon_L$:
\be
{{|\d_a\mu|}\over{\mu}}=4
{{|\d_aT|}\over {T}}<H(12\epsilon_1+
{\t{24\over5}}\epsilon_2)\,,
\ee
\be
{{|\sigma_{ab}|}\over\Theta}<{\t{5\over3}}\epsilon_1+
3\epsilon_2+{\t{3\over7}}\epsilon_3\,,
\ee
\be
{|\omega_{ab}| \over \Theta}< {\t{10\over3}}\epsilon_1+
{\t{2\over15}}\epsilon_2~,
\ee
\be
{|\d_a \rho| \over \Theta} < {\t{9\over2}}H\epsilon_2+
\left({H \over \Omega_M}\right)[60\epsilon_1+134\epsilon_2+
6\epsilon_3]+\left({\Omega_R \over \Omega_M}\right)
H[16\epsilon_1+{\t{61\over3}}\epsilon_2]~,
\ee
\be
{|\d_a \Theta| \over \Theta} < H({\t{205\over3}}\epsilon_1+
8\epsilon_2)+4(2\Omega_R+\Omega_M)H\epsilon_1~,
\ee
\be
{|E_{ab}| \over \Theta} < H({\t{55\over3}}\epsilon_1+{\t{103\over3}}
\epsilon_2+{\t{23\over7}}\epsilon_3)+{\t{4\over45}}
(11\Omega_R+15\Omega_M)H\epsilon_2~,
\ee
\be
{|H_{ab}| \over \Theta} < H({\t{16\over3}}\epsilon_1+{\t{52\over15}}
\epsilon_2+{\t{1\over21}}\epsilon_3)~.
\ee
\newline
If the dipole is neglected, then we can set $\epsilon_1=0$ in
(58--64). Note also that for the $\Omega_R,\Omega_M$ and $H$ on the
right sides of (50--56) and (58--64)
we may use the $O[0]$ values, i.e. the
values they take in the limiting (background) FRW spacetime,
which has non-interacting
dust and isotropic radiation - exact solutions are given in [33].\\

The bounds (58--64) are the main results of the quest for {\it a
direct link from feasible observational limits on the CBR
to limits on the deviations of the universe from FRW since last
scattering (within our past light cone)}. They are based on the
reasonable assumptions {\bf B1--3, C1--2}. These results give
for the first time a direct relation between CBR observational limits
and limits to anisotropy and inhomogeneity, without assuming a
specific evolutionary model, or the curvature index $k$ of the
background. \\

 From these limits we can obtain conservative estimates of
present--time bounds on
the anisotropy and inhomogeneity of the universe. Let
$$
\epsilon \equiv {\rm max}(\epsilon_1,\epsilon_2,\epsilon_3)
$$
denote the
upper limit of currently observed anisotropy in the CBR temperature
variation, and take $(\Omega_R)_0 \ll 1$. Then (58--64) imply
\be
\left({{|\d_a \mu|} \over \mu}\right)_0 <
17 H_0 \epsilon ~,~~
\left({{|\sigma_{ab}|} \over \Theta}\right)_0 < 6 \epsilon ~,~~
\left({{|\omega_{ab}|} \over \Theta}\right)_0 < 4 \epsilon ~.
\ee
\be
\left({{|E_{ab}|}\over{\Theta}}\right)_0 < [{\t{4\over3}}(\Omega _M)_0
+ 56]H_0 \epsilon~,~~\left({|H_{ab}| \over \Theta}\right)_0 <
10 H_0 \epsilon~,
\ee
\be
\left({|\d_a \Theta| \over \Theta}\right)_0 < [4(\Omega_M)_0+77]
H_0\epsilon~,~~
\left({{|\d_a\rho|} \over \rho}\right)_0 <
C(\Omega) H_0 \epsilon~,
\ee
where $C(\Omega) \equiv 5+ 200/(\Omega_M)_0$.
The latter is a relatively weak limit, consistent with inhomogeneities
in the large--scale matter distribution.
If we are willing to assume that $(\Omega_M)_0 \simeq 1$ today,
we get a reasonably tight limit from (67).
However, the observational evidence points towards a range of
values between 0.1 and 0.3 as more plausible [34]. Including the
lowest limits implied by nucleosynthesis, we can represent the
range of possibilities by a table of values:\\

\begin{tabular}{|c|r|r|r|r|}\hline
  & & & & \\
$(\Omega_M)_0$ & 0.02  & 0.1  & 0.3  &  1 \\
  & & & & \\
\hline
  & & & & \\
$C(\Omega)$    & 10005 & 2005 & 672 & 205 \\
  & & & & \\ \hline
\end{tabular} \\ \\

As we go back in time towards last scattering, whatever value it has
today, $\Omega_M$ will rapidly approach 1. \\

We note that if the dipole is neglected, then we can set
$\epsilon_1=0$ in (58--64) to obtain the revised
``dipole--free" estimates:
$$
\left({{|\d_a \mu|} \over \mu}\right)_0 <
5 H_0 \epsilon ~,~~
\left({{|\sigma_{ab}|} \over \Theta}\right)_0 < 4  \epsilon ~,~~
\left({{|\omega_{ab}|} \over \Theta}\right)_0 < \epsilon ~, \eqno(65a)
$$
$$
\left({{|E_{ab}|}\over{\Theta}}\right)_0 < [{\t{4\over3}}(\Omega_M)_0
+38] H_0 \epsilon~,~~
\left({|H_{ab}| \over \Theta}\right)_0 < 4H_0\epsilon~, \eqno(66a)
$$
$$
\left({|\d_a \Theta| \over \Theta}\right)_0 < 4[(\Omega_M)_0+2]
H_0\epsilon~,~~
\left({{|\d_a\rho|} \over \rho}\right)_0 < \left[
{140 \over (\Omega_M)_0}+5\right] H_0 \epsilon~, \eqno(67a)
$$
which are considerably sharper. We recover the EGS result [32] on
setting $\epsilon = 0$ (exact isotropy implies an exact FRW solution).

\section{Almost Flat--FRW Solutions}
As already pointed out, we are unlikely to have direct
observational information about the spatial gradients of the
temperature multipoles.
In section 4 we used the estimate {\bf C1} that the spatial gradients
are not greater than the time derivatives.
Here we consider a more stringent, but apparently not
unreasonable, assumption on the spatial gradients. In effect we extend
to the spatial gradient bounds the assumption {\bf C2} already made
on the time derivative bounds - i.e. we replace assumptions
{\bf C1--2} by
\begin{quotation}
{\bf D}: the bounds on both time and spatial derivatives of
the temperature multipoles are determined via the characteristic scales
(57) of the CBR:
\end{quotation}
\be
\Theta\epsilon^*_L \simeq {\epsilon_L \over t_{R}}~
\Rightarrow~\epsilon^*_L \simeq {\t{1\over3}}\epsilon_L~,~~~
\Theta \epsilon'_L \simeq {\epsilon_L \over d_{R}}~
\Rightarrow~\epsilon'_L < (\epsilon_1+{\t{2\over5}}
\epsilon_2)\epsilon_L\,,
\ee
where we have used (36) and (58).
It immediately follows from (68)
that $\epsilon'_L=O[2]$, and hence by (44) and (41)
\be
\d_aq_b\simeq 0\,,~~~~\d_a\pi_{bc}\simeq 0\,,
{}~~~~\d_a\xi_{bcd}\simeq 0~,~~~\dots
\ee
Thus from an apparently reasonable assumption on the spatial gradients, we
are led to the vanishing at first order of vector and tensor
inhomogeneities in
the radiation. The point is that the radiation time scale $t_R$
is a zero order quantity (it exists for exactly isotropic CBR), while
the radiation length scale $d_R$ is first order - it is only
finite when there are inhomogeneities in the CBR. If
inhomogeneities in the temperature fluctuation $\tau$ are determined
by the characteristic inhomogeneity scale, then, as shown by (68),
they are negligible in comparison with anisotropies.
This in turn leads to a very restrictive condition:\\
\begin{quotation}
{\it if the radiation energy flux and anisotropic stress are
homogeneous to first order, }
\end{quotation}
i.e. if (69) holds, which will follow if {\bf D} is true, then \\
\begin{quotation} either {\it the
spacetime is FRW to first order}, \,or \,{\it the spatial curvature vanishes
to first order and so the background has a flat FRW geometry.}
\end{quotation}

This follows as a special case of the results {\bf (I), (II)} derived
in section 2.3.
Note that it may also be derived (but with greater effort) by
considering the
integrability conditions of the constraint equations (20--24) in the
case that (69) holds.\\

This result may be viewed as providing an alternative motivation for
the almost flat--FRW model of the universe. Furthermore, we are able
to reduce the system of dynamical equations in this model to a pair of
linear evolution equations, for the shear and energy flux.
Using the $O[1]$
identity (26) with (69), the spatial gradients of (13) and (14) imply
\be
\d_a\d_b\mu\simeq 0 ~~
\Rightarrow~~\omega_{ab}\simeq 0~,~~~
\d_a\sigma_{bc}\simeq 0\,,
\ee
where we have used (27). Then (70) reduces (22) and the spatial
gradient of (16) to
\be
H_{ab} \simeq 0 \,,~~~~~ \d_aE_{bc} \simeq 0\,.
\ee

By (69--71), the system of dynamical equations (11--24) closes at
$O[1]$ and reduces to a sub--system for $\sigma_{ab}$, $\pi_{ab}$,
$E_{ab}$
\be
\dot{\sigma}_{ab}+{\t{2\over3}}\Theta\sigma_{ab}+E_{ab}
-{\t{1\over2}}\pi_{ab} \simeq 0\,,
\ee
\be
\dot{\pi}_{ab}+{\t{4\over3}}\Theta\pi_{ab}+
{\t{8\over15}}\mu\sigma_{ab} \simeq 0\,,
\ee
\be
\dot{E}_{ab}+\Theta E_{ab}+({\t{1\over2}}\rho+
{\t{2\over5}}\mu)\sigma_{ab}-{\t{1\over2}}\Theta
\pi_{ab} \simeq 0 \,,
\ee
and a sub--system for $q_a$, $\d_a \psi$ ($\psi
\equiv \mu,\rho,\Theta$):
\be
\dot{q}_a+{\t{4\over3}}\Theta q_a+{\t{1\over3}}
\d_a\mu \simeq 0\,,
\ee
\be
2\d_a\Theta-3q_a \simeq 0 \,,
\ee
\be
\d_a(\rho+\mu)-\Theta q_a \simeq 0\,.
\ee

These sub--systems represent a {\it decoupling of the anisotropy and
inhomogeneity,} since (72--74) contain no spatial gradients (because
$H_{ab}$, $\omega_{ab}$ and the acceleration vanish to the
order of the calculation, this solution belongs to the `silent
universe' class recently examined by Mataresse et al [35,36]). In the
special case where we assume that the temperature dipole is
negligible, (41) and (75--77) show
$$
q_a \simeq 0~~\Rightarrow~~\d_a\mu \simeq
\d_a\rho \simeq \d_a\Theta \simeq 0
$$
which, together with (69--71), shows that the spacetime is homogeneous
to first order.
Thus in the dipole-free case when {\bf D} is assumed, the spacetime
is Bianchi I to the accuracy of the calculation.
Conversely,
\begin{quotation}
{\it if there is inhomogeneity in
the radiation and matter, and if the CBR characteristic length scale
determines the CBR temperature inhomogeneity (i.e. if~{\bf D} holds),
then the dipole cannot be negligible after correction for peculiar
velocities.}
\end{quotation}

 From now on we will assume that the dipole is not negligible, i.e.
$q_a \neq O[2]$. In (72--77), the coefficients $\mu$, $\rho$
and $\Theta$ may be given their FRW zero-order forms, which are
in fact the solutions of (11,12) (using (69)) and (15). In
particular, these equations imply the $O[1]$ Friedmann equation [6]
\be
\Theta^2-3(\rho+\mu) \simeq 0\,.
\ee

Before we consider the decoupling and solving of (72--74)
and (75--77),
we give the limits that they imply, using (46,47) and (68,69):
\be
{{|\d_a\mu|}\over{\mu}}< 12 H \epsilon_1~,~~
{{|\d_a\rho|}\over{\rho}}<16\left({\Omega_R \over
\Omega_M}\right)H\epsilon_1~,~~
{{|\d_a\Theta|}\over{\Theta}}< 2\Omega_R H \epsilon_1~,~~
{{|\sigma_{ab}|}\over{\Theta}}<3\epsilon_2\,.
\ee
Note that a bound on $E_{ab}$ does not follow directly. These results
sharpen the bounds given by (58,59,61,62) (recalling that
{\bf C1--2} have been replaced by {\bf D}). In fact, the bound on the
matter inhomogeneity is drastically sharpened at late times, when
$\Omega_R \ll \Omega_M$. The source of this is the disappearance of
the gradient of the electric Weyl tensor which
produced the $H/\Omega_M$ term of (61). This gradient, along with all
tensor
gradients (i.e. gradients of vectors and tensors), drops out by virtue
of {\bf D}. In other words:
\begin{quotation}
 {\it if only scalar inhomogeneities occur (i.e. all vectors and tensors are
gradients of scalars), then the matter inhomogeneities are rapidly suppressed
at late times.}
\end{quotation}

A potential problem arising from (79) is that the limits suggest
the matter inhomogeneity is much less than the radiation
inhomogeneity at late times.
However, the upper limits do not force this to occur, they simply
allow the possibility. Below we will give an example of a
late--time solution where the matter inhomogeneity is greater than
the radiation inhomogeneity.\\

By taking $u^a\nabla_a$ derivatives and using (15) and (78), the
sub--systems (69--71) and (72--74) may be decoupled:
\be
\ddot{\sigma}^{\displaystyle\cdot}_{ab}+\Theta^{-1}
({\t{10\over3}}\Theta^2+\mu
+{\t{1\over2}}\rho)\ddot{\sigma}_{ab}+({\t{36\over55}}
\mu+{\t{43\over6}}\rho)\dot{\sigma}_{ab}\,+
\ee
$$
+\,\Theta^{-1}({\t{44\over45}}\mu\Theta^2+
{\t{7\over6}}\rho\Theta^2-{\t{44\over15}}\mu^2
-{\t{9\over4}}\rho^2-{\t{26\over5}}\mu\rho)
\sigma_{ab} \simeq 0\,,
$$
\be
\ddot{q}_a+3\Theta \dot{q}_a+({\t{16\over9}}\Theta^2-
2\mu-{\t{2\over3}}\rho)q_a \simeq 0\,.
\ee

In principle we can obtain the $O[1]$ solution for the
almost flat-FRW model after last scattering as follows.
The solutions of (11), (12) and (78) imply
\be
\mu \simeq rS^{-4}\,,~~~~~\rho = mS^{-3}\,,~~~~~\Theta \simeq
S^{-2}[3(r+mS)]^{1/2}\,,~~~~~\dot{r}\simeq 0=\dot{m}
\ee
which allow us to reduce (80,81) to linear ODE's in $S$.
We write
\be
\sigma_{ab}\simeq A_{ab}^{(I)}\Sigma_{(I)}(S)\,,~~~~~~I=1,2,3
\ee
\be
q_a \simeq B_a^{(\Lambda)}Q_{(\Lambda)}(S)\,,~~~~~~\Lambda=1,2
\ee
with $\dot{A}_{ab}^{(I)}\simeq 0 \simeq \dot{B}_a^{(\Lambda)}$. Then
$\Sigma_{(I)}$ are linearly independent solutions of
\be
\Sigma'''+\left[{{2(8r+9mS)} \over S(r+mS)}\right]\Sigma''-
\left[{r(94r+89mS) \over
S^2(r+mS)^2}\right]\Sigma'\,-
\ee
$$
-\,\left[{(480r^2+1006rmS+525m^2S^2)
\over 20S^3(r+mS)^2}\right]\Sigma \simeq 0
$$
and $Q_{(\Lambda)}$ are linearly independent solutions of
\be
Q''+\left[{3(4r+5mS) \over 2S(r+mS)}\right]Q'+
\left[{2(5r+7mS) \over S^2(r+mS)}\right]Q \simeq 0\,.
\ee
\newline

Given the solution $\Sigma(S)$ of (85), we find from (72,73) that
\be
\pi_{ab}\simeq {C_{ab} \over S^4}-\left({8r \over 5S^4}\right)
A_{ab}^{(I)}\int \left({S\Sigma_{(I)} \over [3(r+mS)]^{1/2}}\right)dS\,,
\ee
\be
E_{ab} \simeq {C_{ab} \over 2S^4}-A_{ab}^{(I)}\left[\left({[3(r+mS)]^
{1/2} \over 3S}\right)
\Sigma'_{(I)}+\left({4r \over 5S^4}\right)
\int \left({S\Sigma_{(I)} \over [3(r+mS)]^{1/2}}\right)dS \right]
\ee
where $\dot{C}_{ab} \simeq 0$.\\

Similarly, given the solution $Q(S)$ of (86), we find from
(75--77) that
\be
\d_a\mu \simeq -[3(r+mS)]^{1/2} B_a^{(\Lambda)}
[Q'_{(\Lambda)}+{4 \over S}Q_{(\Lambda)}]\,,
\ee
\be
\d_a\Theta \simeq {\t{3\over2}}B_a^{(\Lambda)}
Q_{(\Lambda)}\,,
\ee
\be
\d_a\rho \simeq -[3(r+mS)]^{1/2} B_a^{(\Lambda)}
[Q'_{(\Lambda)}+{5 \over S}Q_{(\Lambda)}]\,.
\ee
Note that because two time derivatives were needed in the decoupling
that led to (80), there will be a consistency condition imposed on the
integration constants $A_{ab}^{(I)}$, $C_{ab}$ via (74). \\

Thus (78), (83,87,88) and (84,89--91) represent {\it an exact
solution of the linearised equations governing the metric,
CBR and matter after last scattering in a universe subject to the
observational assumption {\bf D}, and with anisotropic radiation}.
The explicit analytic form of the solutions depends on finding
explicitly the solutions to the ODE's (85,86).
We can provide explicit solutions for {\it late times} (i.e. long
after last scattering) when
\be
\mu \ll \rho \simeq {\t{1\over3}}\Theta^2 \ll 1
\ee
using (78). By (92), (80) simplifies to
\be
\ddot{\sigma}^{\displaystyle\cdot}_{ab}+
{\t{7\over2}}\Theta\ddot{\sigma}_{ab}+
{\t{43\over18}}\Theta^2\dot{\sigma}_{ab}+
{\t{5\over36}}\Theta^3\sigma_{ab} \simeq 0\,.
\ee
Writing $\sigma_{ab}\simeq U_{ab}\Theta^n$, where $\dot{U}_{ab}
\simeq 0$, we find from (93) that $n={1\over3}, {5\over3}, 2$. However
$n={1\over3}$ violates (51) at late times. Thus an acceptable
solution to (93) is
\be
\sigma_{ab} \simeq U_{ab}\Theta^{5/3}+V_{ab}\Theta^2\,,~~~~~
\dot{U}_{ab}\simeq 0 \simeq \dot{V}_{ab}\,.
\ee
Using (92) and (94) in (72--74) we find that the consistency
condition on the constants of integration gives $V_{ab}\simeq 0$ and
the late-time solution is
\be
\sigma_{ab}\simeq U_{ab}\Theta^{5/3}\,,~~~~\pi_{ab} \simeq
{\t{1\over6}}U_{ab}\Theta^{8/3}\simeq {\t{1\over6}}
\Theta\sigma_{ab}\,,~~~~E_{ab}\simeq
{\t{1\over4}}U_{ab}\Theta^{8/3}\simeq {\t{3\over2}}
\pi_{ab}\,.
\ee
\newline
Now we use (92) to simplify (81) to
\be
\ddot{q}_a+3\Theta \dot{q}_a+{\t{14\over9}}\Theta^2q_a\simeq 0\,.
\ee
With $q_a\simeq J_a\Theta^n$, $\dot{J}_a\simeq 0$, we get $n={7\over3}
,{8\over3}$, neither of which violates the limit (46). Thus a solution
of (96) is
\be
q_a \simeq J_a\Theta^{7/3}+K_a\Theta^{8/3}\,,~~~~~
\dot{J}_a \simeq 0 \simeq \dot{K}_a \,,
\ee
and this leads via (75--77) to
\be
\d_a\mu \simeq -{\t{1\over2}}J_a\Theta^{10/3}\,,
{}~~~\d_a\Theta \simeq {\t{3\over2}}q_a\,,~~~
\d_a\rho \simeq {\t{3\over2}}J_a\Theta^{10/3}+
K_a\Theta^{11/3}\,.
\ee

Thus we have presented an exact and explicit late--time solution (94,
95,97,98), which in particular provides a quantitative measure for
the rate of approach towards the limiting background solution.
This solution also allows us to tighten the limit in (79) on shear
at late times, and to give a late-time limit on the Weyl tensor:
by (46) and (95) we have
\be
{|\sigma_{ab}| \over \Theta}< \left({16\Omega_R \over 15\Omega_M}
\right)\epsilon_2\,,~~~~~~~~~
{|E_{ab}| \over \Theta}< {\t{4\over5}}\Omega_R H
\epsilon_2\,.
\ee
Furthermore, the solution contains spacetimes in
which the late--time radiation inhomogeneity is negligible in
comparison with the matter inhomogeneity. This arises when
$J_a$ and $K_a$ in (98) are chosen to ensure that
\be
{{|\d_a\mu|} \over \mu} \ll
{{|\d_a\rho|} \over \rho} < 16 \left({\Omega_R \over
\Omega_M}\right)H \epsilon_1
\ee
where the final inequality follows from (79), consistently with
(97,98). The particular choice of $J_a \simeq 0$, i.e. radiation
inhomogeneity vanishing to first order at late times,
will achieve (100), with
\be
\d_a\mu \simeq 0~,~~~~
{{|\d_a\rho|} \over \rho} < 4 \left({\Omega_R \over
\Omega_M}\right)H \epsilon_1~,
\ee
by (97,98) and (46). \\

\section{Conclusion}
We have seen how a series of assumptions of increasing sharpness
(incorporating the inevitable Copernican supposition)
leads to  increasingly powerful deductions from the CBR anisotropy.
As emphasized before, these limits are independent of detailed assumptions
about the dynamical history of matter in the universe, and provide an
alternative mode of analysis to the usual approaches. This analysis
has the advantage of being both covariant and gauge-invariant [11,12].
It gives somewhat less information than the usual approaches based on
the Sachs-Wolfe effect and its generalisations, precisely because it is
more model independent; however this also means that its conclusions are
more robust than those more standard analyses, as they do not depend
so much on the assumptions of particular evolutionary models.
In particular
they are not dependent on whether or not inflation took place,
and whether or
not the density parameter $\Omega$ is near the critical value.\\

The analysis proceeds through a series of increasingly restrictive
Copernican assumptions about the nature of CBR anisotropies in an
open neighbourhood about our world line, which is envisaged as
including the observable region of the universe. Such assumptions are
inevitable if we wish to justify the assumption of an almost-FRW model
[7,8]. The qualitative assumption {\bf A1} (Section 1) is sharpened to
the quantitative assumptions
{\bf B1-3} (Section 2) leading to the limits (50--56),
which can be sharpened
a little if we assume that the matter and radiation frames coincide
(i.e. if there is
no CBR temperature dipole).
Somewhat sharper assumptions {\bf C1--2} (Section 4) give better limits,
leading to our estimates (58--64) and (65-67), which are further
sharpened if the dipole
can be neglected to give (65a,66a,67a). These equations make
quantitative the
results of [6] (the universe is almost FRW, see {\bf A2}) and include as a
special case the Ehlers-Geren-Sachs exact theorem [32]. They confirm
previous estimates based on the CBR anisotropies that the shear and vorticity
are at most about $10^{-3}$ of the expansion (on choosing $\epsilon =
10^{-4}$, to concur with recent CBR anisotropy measurements). \\

We regard these assumptions and limits as highly plausible, and believe
they are useful not only in terms of the limits obtained but also in
making quite explicit the kind of Copernican assumptions one has to make
in order to extract information from the CBR anisotropies (such assumptions
necessarily underlie the standard Sachs-Wolfe type analyses, because these
analyses
assume {\bf A2} as their starting point, but they do so in a somewhat
hidden way). More debatable are the stronger assumptions {\bf D} of
section 5, which seem on the face of it quite plausible but then lead
to very restrictive conclusions: either the universe is FRW to first order
(that is, its difference from a FRW geometry is at most second order) or
it has a flat background FRW geometry (to the accuracy of our first-order
calculation).  When this is true we can get explicit solutions of
the equations representing the combined matter and radiation system - but
they only allow spatial inhomogeneity when the dipole term cannot be
neglected.\\

Thus some may wish to adopt these stronger assumptions, while others may feel
the conclusions are too strong and therefore the assumptions should be
questioned. We are open minded in this matter; the main point is that
the analysis presented here makes quite clear the range of possible
assumptions, and their consequences when we take the Einstein-Liouville
equations (and consequent propagation and conservation equations) into
account. \\

Finally we believe this paper shows well the utility of the covariant harmonic
approach to both perturbations and to kinetic theory. In particular
in examining kinetic effects, it makes quite
clear how only the first three harmonic terms in the distribution function
explicitly enter the field equations; the
anisotropies represented by higher harmonics can affect the geometry only
by {\it cascading} [15]: that is, by inducing anisotropies in the lower order
harmonics through divergence or gradient terms, as in equation (10), or
through collisions [16,37].
\newline
\newline {\bf Acknowledgements:} RM was supported by research grants
from the FRD, Witwatersrand University and Portsmouth University. GE and
WS thank the FRD for financial support.
\newpage
{\bf REFERENCES}
\newline
\newline
1. GF Smoot {\it et al.} (1992) {\it Astrophys J Lett} {\bf 396}, L1\\
2. EL Wright {\it et al.} (1994) {\it Astrophys J} {\bf 420}, 1 \\
3. T Gaier {\it et al.} (1992) {\it Astrophys J Lett} {\bf 398}, L1 \\
4. M Dragovan {\it et al.} (1994) {\it Astrophys J Lett} {\bf 427},
L67 \\
5. RB Partridge (1994) {\it Class Quantum Grav} {\bf 11}, A153 \\
6. WR Stoeger, R Maartens and GFR Ellis (1994) {\it Cape Town
University Preprint} 1994/9 (to appear {\it Astrophys J}) \\
7. G F R Ellis (1975) {\em Qu Journ Roy Ast Soc} {\bf 16}, 245-264 \\
8. D R Matravers, W R Stoeger, G F R Ellis (1994) {\it Portsmouth
University Preprint} 1994/7 (to appear {\it QJRAS})\\
9. GFR Ellis, M Bruni and J Hwang (1990) {\it Phys Rev} {\bf D 42},
1035 \\
10. BACC Bassett, PKS Dunsby and GFR Ellis (1994) {\it Cape Town
University Preprint} 1994/4 \\
11. GFR Ellis and M Bruni (1989) {\it Phys Rev} {\bf D 40}, 1804 \\
12. PKS Dunsby, M Bruni and GFR Ellis (1992) {\it Astrophys J}
{\bf 395}, 54 \\
13. SW Hawking (1966) {\it Astrophys J} {\bf 145}, 544 \\
14. JM Bardeen (1980) {\it Phys Rev} {\bf D 22}, 1882 \\
15. GFR Ellis, DR Matravers and R Treciokas (1983) {\it Ann Phys}
{\bf 150}, 455 \\
16. GFR Ellis, R Treciokas and DR Matravers (1983) {\it Ann Phys}
{\bf 150}, 487 \\
17. R Durrer (1989) {\it Astron Astrophys} {\bf 208}, 1 \\
18. AK Rebhan and DJ Schwarz (1994) {\it Preprint GR-QC} 9403032 \\
19. T Padmanabhan (1993) {\it Structure Formation in the Universe}
(Cambridge) \\
20. S Dodelson, L Knox and EW Kolb (1994) {\it Phys Rev Lett} {\bf
72}, 3444 \\
21. R Crittenden, J Bond, R Davis, G Efstathiou and P Steinhardt
(1993) {\it Phys Rev Lett} {\bf 71}, 324 \\
22. EJ Copeland (1993) {\it Contemp Phys} {\bf 34}, 303 \\
23. GFR Ellis (1971) in {\it General Relativity and Cosmology - Proc.
47th E Fermi Summer School} ed RK Sachs (Academic) \\
24. EJ Copeland, EW Kolb, AR Liddle and JE Lidsey (1993) {\it Phys
Rev} {\bf D 48}, 2529 \\
25. PJE Peebles and JT Yu (1970) {\it Astrophys J} {\bf 162}, 815 \\
26. ML Wilson and J Silk (1981) {\it Astrophys J} {\bf 243}, 14 \\
27. W Hu, D Scott and J Silk (1994) {\it Phys Rev} {\bf D 49}, 648 \\
28. A Bonnano and V Romano (1994) {\it Phys Rev} {\bf D 49}, 6450 \\
29. JR Bond and G Efstathiou (1987) {\it Mon Not Roy Ast Soc}
{\bf 226}, 655 \\
30. X Luo (1994) {\it Astrophys J Lett} {\bf 427}, L71 \\
31. S Torres, R Fabbri and R Ruffini (1994) {\it Astron Astrophys}
{\bf 287}, 15 \\
32. J Ehlers, P Geren, R K Sachs (1968) {\it J Maths Phys} {\bf 9}, 1344\\
33. GFR Ellis (1987) in {\it Vth Brazilian School of Cosmology and
Gravitation} ed M Novello (World Scientific)\\
34. P Coles and GFR Ellis (1994) {\it Nature} August 25th\\
35. S Mataresse, O Pantano, D Saez (1993) {\it Phys Rev} {\bf D47},
1311 \\
36. M Bruni, S Matarresse, O Pantano (1994), to appear, {\it Astrophys J}\\
37. D R Matravers and G F R Ellis: {\em Class and Quantum Gravity} {\bf
6}, 369-381 (1989); {\bf 7}, 1869-1875 (1990).\\


\end{document}